\documentclass[12pt,preprint]{aastex}
\usepackage{underscore}
\usepackage{color}

\slugcomment{ Manuscript Accepted on 2017 September 5}

\shorttitle{Polarimetry of (1566) Icarus} 
\shortauthors{Ishiguro et al.}

\begin{document}

\title{Polarimetric Study of Near-Earth Asteroid (1566) Icarus}

\author{Masateru ISHIGURO}
\affil{Department of Physics and Astronomy, Seoul National University,
Gwanak, Seoul 151-742, South Korea}

\author{Daisuke KURODA}
\affil{Okayama Astrophysical Observatory, National Astronomical
Observatory of Japan, National Institutes of Natural Science, Asakuchi, Okayama 719-0232, Japan}

\author{Makoto WATANABE}
\affil{Department of Applied Physics, Okayama University of Science, 1-1 Ridai-cho,
Kita-ku, Okayama, Okayama 700-0005, Japan}

\author{Yoonsoo P. BACH, Jooyeon KIM, Mingyeong LEE}
\affil{Department of Physics and Astronomy, Seoul National University,
Gwanak, Seoul 151-742, South Korea}

\author{Tomohiko SEKIGUCHI}
\affil{Asahikawa Campus, Hokkaido University of Education, 9 Hokumon,
Asahikawa 070-8621, Japan}

\author{Hiroyuki NAITO}
\affil{Nayoro Observatory, 157-1 Nisshin, Nayoro City, Hokkaido 096-0066, Japan}

\author{Katsuhito OHTSUKA}
\affil{National Astronomical Observatory of Japan, Osawa 2-21-1, Mitaka, Tokyo 181-8588, Japan}

\author{Hidekazu HANAYAMA}
\affil{Ishigakijima Astronomical Observatory, National Astronomical
Observatory of Japan, Ishigaki, Okinawa 907-0024, Japan}

\author{Sunao HASEGAWA}
\affil{Institute of Space and Astronautical Science (ISAS),
Japan Aerospace Exploration Agency (JAXA), Sagamihara, Kanagawa 252-5210, Japan}

\author{Fumihiko USUI}
\affil{Center for Planetary Science,
Graduate School of Science, Kobe University,
7-1-48, Minatojima-Minamimachi, Chuo-Ku, Kobe 650-0047, Japan}

\author{Seitaro URAKAWA}
\affil{Bisei Spaceguard Center, Japan Spaceguard Association, 1716-3 Okura, Bisei-cho,
Ibara, Okayama 714-1411, Japan}

\author{Masataka IMAI, Mitsuteru SATO, Kiyoshi KURAMOTO}
\affil{Department of Cosmosciences, Graduate School of Science, Hokkaido University, Kita-ku, Sapporo, Hokkaido 060-0810, Japan}

\begin{abstract}
We conducted a polarimetric observation of the fast--rotating near--Earth asteroid (1566) Icarus at large phase (Sun--asteroid--observer's) angles $\alpha$= 57\arcdeg--141\arcdeg\ around the 2015 summer solstice. We found that the maximum values of the linear polarization degree are $P_\mathrm{max}$=7.32$\pm$0.25 \% at  phase angles of $\alpha_\mathrm{max}$=124\arcdeg$\pm$8\arcdeg\ in the $V$-band and $P_\mathrm{max}$=7.04$\pm$0.21 \% at $\alpha_\mathrm{max}$=124\arcdeg$\pm$6\arcdeg\ in the $R_\mathrm{C}$--band. Applying the polarimetric slope--albedo empirical law, we derived a geometric albedo of $p_\mathrm{V}$=0.25$\pm$0.02, which is in agreement with that of Q-type taxonomic asteroids. $\alpha_\mathrm{max}$ is unambiguously larger than that of Mercury, the Moon, and another near--Earth S--type asteroid (4179) Toutatis but consistent with laboratory samples with hundreds of microns in size. The combination of the maximum polarization degree and the geometric albedo is in accordance with terrestrial rocks with a diameter of several hundreds of micrometers. The photometric function indicates a large macroscopic roughness.  We hypothesize that the unique environment (i.e., the small perihelion distance $q$=0.187 au and a short rotational period of $T_\mathrm{rot}$=2.27 hours) may be attributed to the paucity of small grains on the surface, as indicated on (3200) Phaethon.
\end{abstract}


\section{Introduction}
\label{sec:introduction}

The polarimetry of solar system airless bodies (e.g., the Moon, Mercury and asteroids) is a useful diagnostic measure for investigating their surface physical properties, such as the albedo and regolith size. The linear polarization degree $P_\mathrm{r}$ is defined as

\begin{eqnarray}
P_\mathrm{r}=\frac{I_\bot-I_\|}{I_\bot+I_\|},
\label{eq:P}
\end{eqnarray}

\noindent where $I_\bot$ and $I_\|$ denote the intensities of scattered light measured with respect to the scattering plane. In general, $P_\mathrm{r}$ exhibits a strong dependence on the phase angle (Sun--target--observer's angle, $\alpha$), consisting of a negative branch at $\alpha\lesssim$20\arcdeg\ with a minimum value $P_\mathrm{min}$ at the phase angle $\alpha_\mathrm{min} \sim$10\arcdeg\ and a positive branch at $\alpha \gtrsim$20\arcdeg\ with a maximum value $P_\mathrm{max}$
at $\alpha_\mathrm{max}\sim$100\arcdeg\ \citep[see, e.g.,][]{Geake1986}.

It is well known that the albedo (the $V$-band geometric albedo, $p_\mathrm{V}$, is often referenced) of objects has a strong correlation with the polarization degree (so-called Umov's law) because multiple scattering (which is dependent on the single--scattering albedo) randomizes polarization vectors and eventually weakens the polarization degree of the scattered signal from the bodies. Moreover, it was noted that $P_\mathrm{max}$ and $\alpha_\mathrm{max}$ have a moderate correlation with grain size \citep{Bowell1972,Dollfus1998,Shkuratov1992}. While intensive polarimetric research on asteroids has been conducted at small phase angles \citep[$\alpha \lesssim$40\arcdeg,][]{Belskaya2009,GH2011,GH2012,GH2014,Cellino2016,Belskaya2017}, polarimetric studies of asteroids at large $\alpha$ values remain less common, most likely because of fewer opportunities (limited to near-Earth asteroids, NEAs) and observational difficulty (small solar elongation).

Here, we would like to stress the superiority of observatories at middle latitudes ($|l|\sim$45\arcdeg) for NEA observations at large $\alpha$ values. During the summer solstice, the Sun does not set at latitudes of $|l|>$66.6\arcdeg\ (a phenomenon called the midnight Sun); in addition, astronomical twilight lasts through the night at observatories at $|l|>$48.6\arcdeg, making it difficult to make astronomical observations. When we conduct observations at observatories at longitudes slightly lower than $|l|$=48.6\arcdeg, we are able to observe NEAs around the Sun as if we were using the Earth as a coronagraph. Taking advantage of this location, we conducted a polarimetric observation of an NEA, (1566) Icarus (=1949 MA), at the Nayoro Observatory ($l$=+44.4\arcdeg) in Hokkaido, Japan, around the summer solstice in 2015. Icarus is one of the Apollo asteroids that has a small perihelion distance of $q$=0.187 au. Such asteroids with a small perihelion distance have gained the attention of solar system scientists interested in understanding the mass erosion mechanisms on these bodies \citep{Jewitt2013b,Granvik2016}.  Icarus has a diameter of 0.8--1.3 km \citep{Veeder1989,Mahapatra1999,Harris2002,Nugent2015} and a short rotational period of 2.27 hours \citep{Miner1969,Gehrels1970,Harris1998,Warner2015}. The asteroid is classified as an S--type asteroid \citep{Chapman1975} or, more specifically, as a Q--type  asteroid \citep{DeMeo2014}. Thanks to the favorable location of the observatory, we were able to acquire polarimetric data up to $\alpha$=141\arcdeg\ and imaging data up to $\alpha$=145\arcdeg. The phase angle is overwhelmingly larger than those of previous polarimetric observations at large phase angles \citep*[i.e., $\alpha<$106\arcdeg\ for (1685) Toro, 115\arcdeg\ for (23187) 2000 PN$_9$, and 118\arcdeg\ for (4179) Toutatis,][]{Kiselev1990,Belskaya2009,Ishiguro1997}. We describe our observations in Section 2, the data reduction in Section 3, and the results in Section 4. Finally, we compare our polarimetric results with those of laboratory samples and solar system airless bodies, and  we consider the surface regolith properties of the asteroid in Section 5.

\clearpage
\section{Observation}
\label{sec:observation}

The journal of our observations is given in Table \ref{tab:t1}.

\begin{figure}
\epsscale{1.0}
\plotone{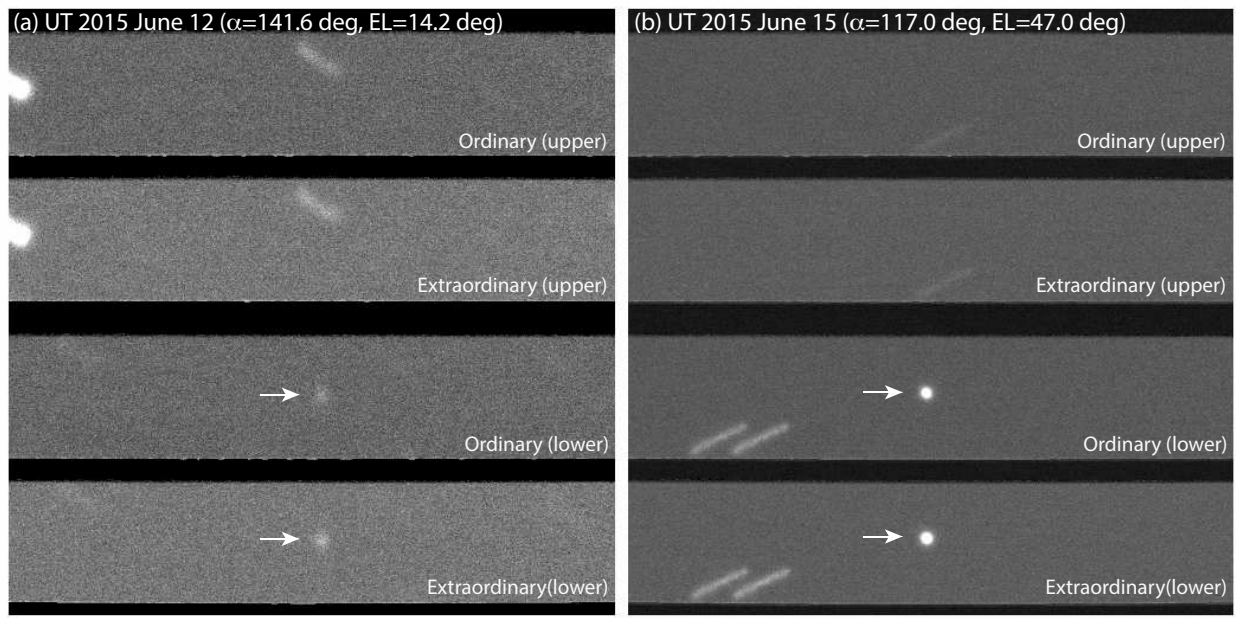}
\caption{Example snapshot $R_\mathrm{C}$--band images in the polarization mode of the MSI captured at (a) 13:24 on UT 2015 June 12
($\alpha$=141.6\arcdeg and an elevation of 14.2\arcdeg) and (b) 11:50 on UT 2015 June 15 ($\alpha$=117.0\arcdeg and an elevation of 47.0\arcdeg) with exposure times of 60 sec and 30 sec, respectively.
The MSI FOV is divided into two areas of the sky using a slit mask for the polarimetry, and each area is split into two components consisting of
ordinary rays and extraordinary rays in these images. Since the asteroid was put in the center of the lower (i.e., southern) slit and tracked
in an asteroid-tracking mode of the telescope, light from the asteroid appears as two point-like objects on the MSI detector (indicated by arrows).
The extended objects are field stars that are stretched out by the asteroid tracking mode. The FOV of each tile is 3.3\arcmin $\times$ 0.7\arcmin.
Each image has a standard orientation in the sky; that is, north is up, and east is to the left.
\label{fig:image}}
\end{figure}

We conducted observations for eight nights from UT 2015 June 11 to UT 2015 June 20 using the 1.6 m Pirka telescope at the Nayoro Observatory (142\arcdeg28$'$$58.0''$, +44\arcdeg22$'$25.1$''$, 192.1 m, observatory code number Q33). The observatory has been operated since 2011 by the Faculty of Science, Hokkaido University, Japan. We utilized a Multi-Spectral Imager (MSI) mounted at the $f/$12 Cassegrain focus of the Pirka telescope. The combination of the telescope and the instrument enables the acquisition of images that cover a 3.3\arcmin $\times$ 3.3\arcmin\ field-of-view (FOV) with a 0.39\arcsec\ pixel resolution \citep{Watanabe2012}. We conducted an imaging observation on the first night, June 11, to test the non-sidereal tracking of the fast moving object at a low elevation ($\sim$10\arcdeg). These data were used for the study of photometric functions presented in Section \ref{subsec:photometry}. After the second night, we made an imaging polarimetric observation from June 12--20 using a polarimetric module comprising a Wollaston prism and a half-wave plate. To avoid the blending of ordinary and extraordinary rays, a two-slit mask was placed at the focal plane for the polarimetric observation. With the mask, the FOV was subdivided into two adjacent sky areas of 3.3\arcmin $\times$ 0.7\arcmin\ each and separated by 1.7\arcmin\ (see Figure \ref{fig:image}). We chose standard Johnson-Cousins $V$ and $R_\mathrm{C}$--band filters for this study to examine the wavelength dependence of the polarization degree.

We took polarimetric data with exposure times of either 30 seconds or 60 seconds (depending on the apparent magnitude of the asteroid) for a single frame. Between exposures, the half-wave plate was routinely rotated from 0\arcdeg\ to 45\arcdeg, from 45\arcdeg\ to 22.5\arcdeg, and from 22.5\arcdeg\ to 67.5\arcdeg\ in sequence to acquire one subset of polarimetric data. Once we acquired the subset of data, the pointing direction of the telescope was shifted by +10\arcsec\ and -10\arcsec\ in turn along the east--west axis (the longer axis of the polarization mask) to acquire the other two subsets. This technique (called dithering) can reduce the effects of pixel-to-pixel inhomogeneity that were not substantially corrected by flat-field correction. Accordingly, each set of data consists of twelve exposures (four exposures with different half-wave plate angles $\times$ three locations on the CCD chip with the $\pm$10\arcsec-dithering mode). We took bias frames  before and after the asteroid observations in approximately 3 hour intervals. At the end of nightly observation, we obtained dome flat field data at the same focal position of the telescope as that with which we observed the asteroid.

\clearpage

\section{DATA ANALYSIS}
\label{sec:dataanalysis}
The observed data were analyzed in the same manner as in \citet{Kuroda2015} and \citet{Itoh2016}. The raw observational data were preprocessed using flat images and bias frames by the MSI data reduction package (MSIRED). Cosmic rays on the images were erased using the L.A.Cosmic tool \citep{Dokkum2001}. After these processes, we extracted source fluxes on ordinary and extraordinary parts of the images (see Figure 1) using the aperture photometry package in IRAF. The obtained fluxes were used to derive the Stokes parameters after completing the necessary procedures for the Pirka/MSI data  (see Appendix A). These procedures contain corrections for polarization efficiencies, the subtraction of instrumental polarization, and the conversion into the standard celestial coordinate system.

The linear polarization degree ($P$) and the position angle of polarization ($\theta_\mathrm{P}$) were derived with the following equations:

\begin{eqnarray}
P = \sqrt{\left(q'''_\mathrm{pol}\right)^2+{\left(u'''_\mathrm{pol}\right)}^2}~,
\label{eq:P2}
\end{eqnarray}

\noindent and

\begin{eqnarray}
\theta_\mathrm{P}=\frac{1}{2}\tan^{-1}\left(\frac{u'''_\mathrm{pol}}{q'''_\mathrm{pol}}\right)~,
\label{eq:theta}
\end{eqnarray}

\noindent where $q'''_\mathrm{pol}$ and $u'''_\mathrm{pol}$ are the Stokes parameters $Q$ and $U$, respectively, normalized by $I$ after correcting for instrumental effects. We derived the linear polarization degree with respect to the scattering plane:

\begin{eqnarray}
P_\mathrm{r}=P \cos\left(2\theta_\mathrm{r}\right)~,
\label{eq:Pr}
\end{eqnarray}

\noindent where $\theta_\mathrm{r}$ is given by

\begin{eqnarray}
\theta_\mathrm{r}=\theta_\mathrm{P}-\left(\phi\pm90\arcdeg\right)~,
\label{eq:thetar}
\end{eqnarray}

\noindent where $\phi$ is the position angle of the scattering plane on the sky.

The polarization degree of each set of four exposures has an error of 0.2--5 \% (depending largely on the apparent magnitudes of the nights). We combined these sets of ($q'''$, $u'''$) values to obtain nightly averaged $q'''$ and $u'''$ values, addressing systematic noise ($\delta_{q'''}$ and $\delta_{u'''}$) and random noise ($\sigma_{q'''}$ and $\sigma_{u'''}$), separately. Regarding systematic noise, we took the arithmetic averages ($\delta_{\overline{q}'''}$ and $\delta_{\overline{u}'''}$). Regarding the synthesized random errors, we calculated the variances of the weighted means, given by

\begin{eqnarray}
\sigma^2_{\overline{q}'''}=\frac{1}{\sum^n_{i=1}\left(\sigma_{q_i'''}\right)^{-2}}~~, ~~~~~~~~~
\sigma^2_{\overline{u}'''}=\frac{1}{\sum^n_{i=1}\left(\sigma_{u_i'''}\right)^{-2}}~~,
\label{eq:variance}
\end{eqnarray}

\noindent where $\sigma^2_{\overline{q}'''}$ and $\sigma^2_{\overline{u}'''}$ are the synthesized random errors for $\overline{q}'''$ and $\overline{u}''''$, respectively.
The resultant values for $q'''$ and $u'''$ are given by

\begin{eqnarray}
\overline{q}'''= \sigma^2_{\overline{q}'''} \sum_{i=1}^n \frac{q_i'''}{\sigma^2_{q_i'''}}~~, ~~~~~~~~~
\overline{u}'''= \sigma^2_{\overline{u}'''} \sum_{i=1}^n \frac{u_i'''}{\sigma^2_{u_i'''}}~~,
\label{eq:qufinal}
\end{eqnarray}

\noindent with total errors of

\begin{eqnarray}
\epsilon_{\overline{q}'''}=\sqrt{\sigma^2_{\overline{q}'''}+\delta^2_{\overline{q}'''}}~~, ~~~~~~~~~
\epsilon_{\overline{u}'''}=\sqrt{\sigma^2_{\overline{u}'''}+\delta^2_{\overline{u}'''}}~~.
\label{eq:querror}
\end{eqnarray}

Similarly, using Eqs. (\ref{eq:P})--(\ref{eq:thetar}), we obtained synthesized $P_\mathrm{r}$ and $\theta_\mathrm{P}$ values, as shown in the following sections.

\clearpage
\section{RESULTS}
\label{sec:results}
We summarize our polarimetric results in Table \ref{polresult}. We describe our findings below.

\subsection{Phase Angle Dependence and Polarimetric Color}
\label{subsec:phase}
\begin{figure}
\epsscale{1.0}
\plotone{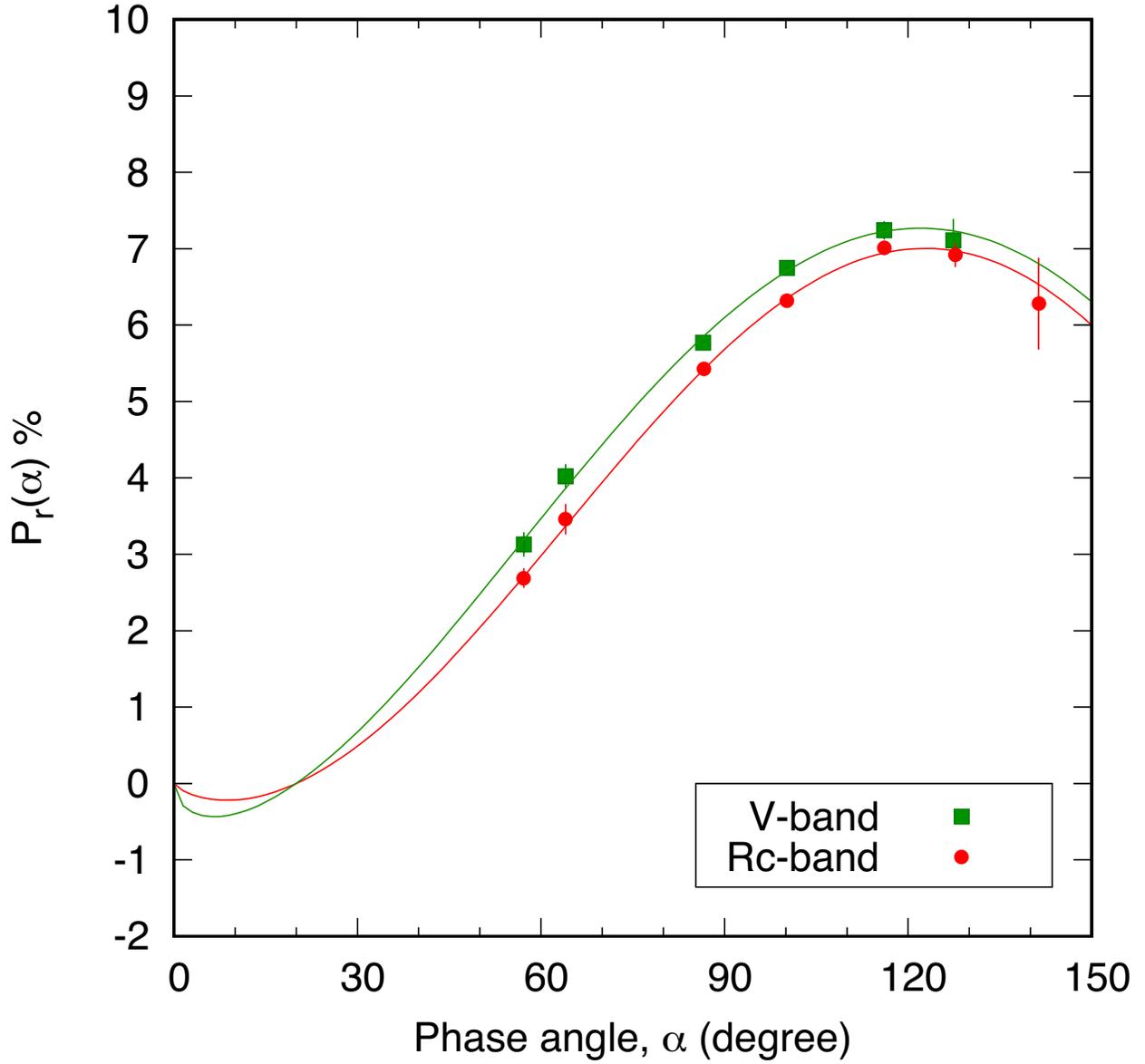}
\caption{Phase angle dependence of polarization degree in the $V$-band and the $R_\mathrm{C}$-band. For reference, we show the fitting lines using Eq. (\ref{eq:LM}) constraining the inversion angle $\alpha_0$=20\arcdeg.
\label{fig:f2}}
\end{figure}

Figure \ref{fig:f2} shows the nightly averaged polarization degrees with respect to the phase angles.
At first glance, we noted that the polarization degree increases almost linearly with increasing phase angle at $\alpha$=60\arcdeg--100\arcdeg, has a peak $\alpha\sim120$\arcdeg, and drops at $\alpha$=125\arcdeg--140\arcdeg.
We also found that the $V$ polarization degrees are higher than the $R_\mathrm{C}$ polarization degrees regardless of the observed phase angles. The average difference is $\Delta P_\mathrm{r}$=$-0.37$$\pm$0.04\%. This trend is similar to that observed for other S-type asteroids \citep{Mukai1997,Lupishko1995,Belskaya2009} and a Q-type asteroid \citep{Fornasier2015}. 

To obtain $\alpha_\mathrm{max}$ and $P_\mathrm{max}$, we fit our data using the Lumme and Muinonen function \citep{Goidet-Devel1995,Penttila2005}:

\begin{eqnarray}
P_\mathrm{r}=b \sin^{c_1} \left(\alpha\right) \cos^{c_2}\left(\frac{\alpha}{2}\right) \sin\left(\alpha-\alpha_0\right),
 \label{eq:LM}
\end{eqnarray}

\noindent where $b$, $c_{1}$, $c_{2}$ and $\alpha_0$ are parameters for fitting. Since our data are not covered at lower phase angles, we fixed the inversion angle $\alpha_0$=20\arcdeg\ (a typical value for S-type and Q-type asteroids, \cite{Belskaya2017}) and derived the other three parameters by weighting with the square of the errors. We obtained $\alpha_\mathrm{max}$=124$\pm$8\arcdeg\ and  $P_\mathrm{max}$=7.32$\pm$0.25\% in the $V$-band and $\alpha_\mathrm{max}$=124$\pm$6\arcdeg\ and $P_\mathrm{max}$=7.04$\pm$0.21\% in the $R_\mathrm{C}$-band. However, we noted that $c_2$ has a negative value, which does not make sense per the original definition of the trigonometric function \citep{Penttila2005}. We discuss this insufficiency and describe the error analysis in Section \ref{sec:discussion}.

\subsection{Rotational Variation in $P_\mathrm{r}$}
\label{subsec:rotation}
\begin{figure}
\epsscale{1.0}
\plotone{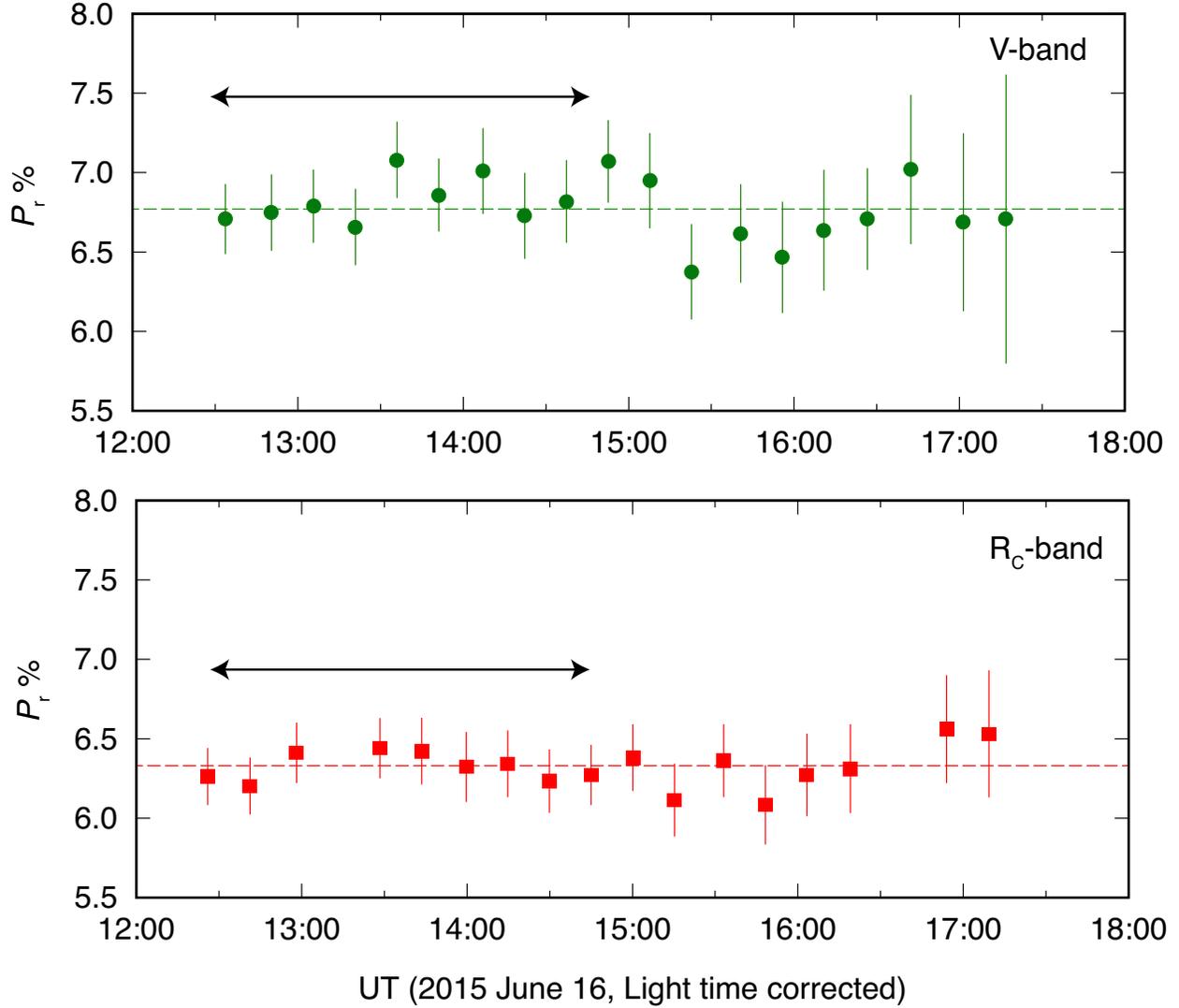}
\caption{Time--dependence of polarization degree in the $V$-band (top) and the $R_\mathrm{C}$-band (bottom) using data taken on UT 2015 June 16 ($\alpha$=100.2\arcdeg). For reference, we show averaged values (dashed lines). The length of arrows corresponds to one rotational period.
\label{fig:phase}}
\end{figure}

Figure \ref{fig:phase} shows the polarization degrees with respect to time from UT 2015 June 16 ($\alpha$=100.2\arcdeg). We choose the data from this night not only because the sky was clear and stable but also because the time coverage was long enough to see a rotational variability in the polarization degree. We combined each set of data taken at three different positions on the detector (see Section \ref{sec:observation}), excluding several images where field stars overlapped the asteroid. The data cover approximately two rotational periods of the asteroid. From this, we found that the polarization degree was notably constant over the quadrant ($\sim$1/4 because $\alpha$$\sim$90\arcdeg) of the surface. We determined the upper limit of the rotational variation in $P_\mathrm{r}$ as 0.3\% in the $V$ band and 0.2 \% in the $R_\mathrm{C}$ band with a 1-sigma confidence level.

It has been reported that some large asteroids show rotational variations of $P_\mathrm{r}$. A notable example is (4) Vesta \citep{Dollfus1989}, which showed a 0.1\% polarimetric variation, and the maximum of the polarization coincides with the lightcurve minimum, suggesting that albedo variation exists on the surface per the controlled polarization degree and visible magnitude. Similarly, (3) Juno, (9) Metis, and (216) Kleopatra showed rotational variations of 0.15--0.27 \%,  $\sim$0.1 \%, and $\sim$0.2 \%, respectively  \citep{Takahashi2009, Nakayama2000,Takahashi2004}. Although the measurement accuracy is too limited to detect such a small variations in the polarization degree, we suggest that Icarus has a quite homogeneous albedo in contrast with these asteroids because our measurement was made at a large phase angle, while these previous detections were made at a small phase angle where the polarization degree itself has small values (1/6$\sim$1/10 of Icarus's $|P_\mathrm{r}|$, i.e., 0.5$\lesssim|P_\mathrm{r}|\lesssim$1.0\%). We will discuss this homogeneity in Section \ref{sec:discussion}.

 \subsection{Photometric Function and Macroscopic Roughness}
 \label{subsec:photometry}
\begin{figure}
\epsscale{1.0}
\plotone{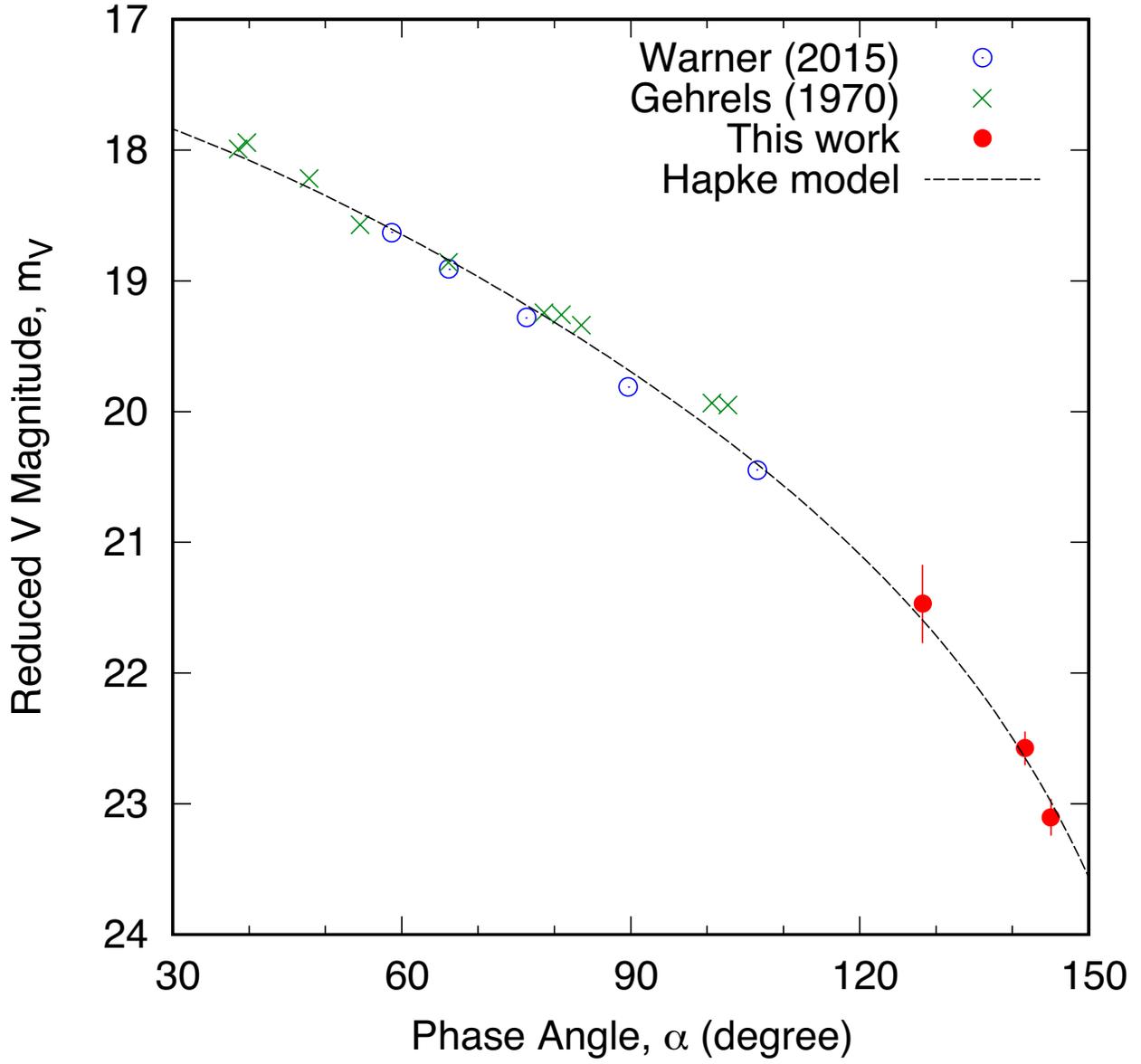}
\caption{Phase angle dependence of the $V$ magnitude. Low-phase data were cited from \citet{Gehrels1970} and \citet{Warner2015}.
\label{fig:mag}}
\end{figure}

 As a byproduct of our polarimetric observation, we took images without using the polarimetric module. These images were obtained mostly in the $R_\mathrm{C}$-band filter when we tested the non-sidereal tracking of the telescope or set the position of  the asteroid in the narrow FOV of the polarization mask. Through comparison with the fluxes of field stars with magnitudes listed in the third U.S. Naval Observatory CCD Astrograph Catalog (UCAC3)  \citep{Zacharias2009}, we derived the $R_\mathrm{C}$-band magnitude of Icarus. Applying the $V-R_\mathrm{C}$ color index of 0.57$\pm$0.08 \citep{Gehrels1970}, the magnitude was converted into the $V$-magnitude. The observed magnitude, $m_\mathrm{V}$, was converted into the reduced $V$-magnitude, $m_\mathrm{V}(1,1,\alpha)$, a magnitude at unit heliocentric and observer's distances that is given by 

\begin{eqnarray}
m_\mathrm{V}(1,1,\alpha)=m_\mathrm{V} - 5~\log(r_\mathrm{h} \Delta)~~,
\label{eq:eq1}
\end{eqnarray}

\noindent where $r_\mathrm{h}$ and $\Delta$ are the heliocentric and observer's distances in au. Figure \ref{fig:mag} is the reduced $V$-magnitude with respect to the phase angle.
In the figure, the magnitude data at $\alpha>$120\arcdeg\ were obtained by us, while the data at $\alpha<$110\arcdeg\ were obtained from \citet{Gehrels1970} and \citet{Warner2015}. 

The phase curve was fitted with the disk-integrated Hapke model \citep{Hapke1993}.
$m_\mathrm{V}(1,1,\alpha)$ data are converted into the logarithm of $I/F$ (where $F$ is the incidence solar irradiance divided by $\pi$, and $I$ is the intensity of
reflected light from the asteroid surface) as

\begin{eqnarray}
-2.5 \log \left(\frac{I}{F}\right) = m_\mathrm{V}(1,1,\alpha) - m_{V \odot} -\frac{5}{2} \log
\left( \frac{\pi}{S}\right) + m_c ~~,
\label{eq:hapke1}
\end{eqnarray}

\noindent
where $m_{V \odot}$=$-26.74$ \citep{Allen1973} is the solar magnitude at 1 au,
$S$ is the geometrical cross section of the asteroid in m$^2$, and
$m_c$ = $-5\log (1.4960 \times 10^{11})$ = $-55.87$ is a constant
to adjust the length unit.  The disk-integrated Hapke function is given by

\begin{eqnarray}
\nonumber
\frac{I}{F} =
 \left[\left(\frac{w}{8}\left[\left(1+B(\alpha)\right)P(\alpha)-1\right]+\frac{r_0}{2}(1-r_0)\right)
  \left(1-\sin\left(\frac{\alpha}{2}\right)\tan\left(\frac{\alpha}{2}\right)
   \ln \left[\cot\left(\frac{\alpha}{4}\right)\right]\right) \right.\\
   \left. +\frac{2}{3}r_0^2
   \left(\frac{\sin(\alpha)+(\pi-\alpha)\cos(\alpha)}{\pi}\right)\right]K(\alpha,\bar{\theta})~~,
\label{eq:hapke2}
\end{eqnarray}

\noindent
where $w$ is the single-particle scattering albedo. $K(\alpha,\bar{\theta})$ is a function that characterizes the surface roughness parameterized by $\bar{\theta}$ \citep{Hapke1984}.
The term $r_0$ is given by

\begin{eqnarray}
r_0 = \frac{1-\sqrt{1-w}}{1+\sqrt{1-w}}~~.
\label{eq:hapke3}
\end{eqnarray}

The opposition effect term $B(\alpha)$ is given by

\begin{eqnarray}
B(\alpha) = \frac{B_0}{1+\frac{\tan(\alpha / 2)}{h}}~~,
\label{eq:hapke4}
\end{eqnarray}

\noindent
where $B_0$ denotes the amplitude of the opposition effect, and $h$
characterizes the width of the opposition effect.

Two parameters of a double Henyey-Greenstein function, $P(\alpha)$ \citep[see, e.g.,][]{Lederer2008}, was employed:

\begin{eqnarray}
P(\alpha)=\frac{(1-c_\mathrm{HG})(1-b_\mathrm{HG}^2)}{(1-2~b_\mathrm{HG} \cos (\alpha) +b_\mathrm{HG}^2)^{3/2}}+\frac{c_\mathrm{HG}(1-b_\mathrm{HG}^2)}{(1+2~b_\mathrm{HG} \cos (\alpha) +b_\mathrm{HG}^2)^{3/2}}~~.
\label{eq:hapke5}
\end{eqnarray}

For the fitting, we fixed opposition parameters as $B_0$=0.02 and $h$=0.141 \citep[from (25143) Itokawa's values,][]{Lederer2008} as an analog of S/Q--type asteroid. By changing the initial values of $b_\mathrm{HG}$, $c_\mathrm{HG}$, $w$, and $\bar{\theta}$ to the range of $0.01 \leq b_\mathrm{HG} \leq 0.8$, $0.01 \leq c \leq 0.8$, $0.1\leq w \leq0.8$, and $5\arcdeg \leq \bar{\theta} \leq 55\arcdeg$, we searched for the best-fit parameters.
From the fitting, we obtained $b_\mathrm{HG}$=0.42$\pm$0.08, $c_\mathrm{HG}$=0.41$\pm$0.20, w=0.48$\pm$0.10, and $\bar{\theta}$=48\arcdeg$\pm$6\arcdeg. Although there are large uncertainties in $b_\mathrm{HG}$ and $c_\mathrm{HG}$, we found that $\bar{\theta}$ is significantly larger for the 10--30 km-sized S-type asteroids (243) Ida and (951) Gaspra \citep{Helfenstein1994,Helfenstein1996}. Note that we assumed the diameter of the asteroid to be 1440 m. If we change the assumed size, $w$ would be different, while $\bar{\theta}$ is nearly constant. Considering the large uncertainty ($\sim$18\%) in the size \citep{Greenberg2017}, the fitting provides a reliable result only for $\bar{\theta}$. The large value of $\bar{\theta}$ may suggest that there are few small particles equivalent to the wavelength (i.e., micrometer or smaller), resulting in the large macroscopic roughness. 
 
\clearpage
\section{Discussion}
\label{sec:discussion}


\subsection{Description for Deriving $P_\mathrm{max}$ and $\alpha_\mathrm{max}$ and Their Errors}
Lumme \& Muinonen's equation, Eq. (\ref{eq:LM}), has been widely used for the fitting of polarimetric phase curves because it produces several key features of the phase curve, including $P_\mathrm{r}$=0 \% at $\alpha$=0\arcdeg, $\alpha_0$ and 180\arcdeg, a negative branch at 0\arcdeg$<\alpha<$$\alpha_0$, and a positive branch at $\alpha>$$\alpha_0$. By definition, the power components $c_1$ and $c_2$ should be positive. The equation has fitted the observed polarimetric phase curves with lower phase angle data well in previous studies \citep[e.g.,][]{Penttila2005}. However, we noted an incompleteness in this function at large phase angles, where we made our observation. We found that the function cannot fit polarimetric data when the phase curve has $\alpha_\mathrm{max}\gtrsim$110 \arcdeg. Thus, the derivative $dP(\alpha)/d\alpha$=0 has no root at $\alpha\gtrsim$110\arcdeg\ in the case of $c_2>0$. 
Our polarimetric data show $dP(\alpha)/d\alpha\sim$0 around $\alpha$=120\arcdeg, indicating $\alpha_\mathrm{max}\sim$120 \arcdeg.  As we mentioned in Section \ref{subsec:phase}, we fit the observed phase profiles with Lumme \& Muinonen's equation without any restriction for ranges of $b$, $c_1$ and $c_2$ and obtained a negative value for $c_2$. Although $c_2$ is out of the range of the original definition, it gives a reasonable fit to the data, as shown in Figure \ref{polresult}. We also noted that the best-fit parameters ($b$, $c_1$, and $c_2$) can be changed for different initial assumptions for $\alpha_0$; however, $\alpha_\mathrm{max}$ and $P_\mathrm{max}$ still remain nearly constant, most likely because our observation covered the range of the maximum where $dP(\alpha)/d\alpha$=0, providing a strict condition for determining reliable estimates for these parameters.

We derived the errors for $\alpha_\mathrm{max}$ and $P_\mathrm{max}$ in the following manner:
We initially obtained $b$=0.0482, $c_1$=0.6521, $c_2$=$-0.7756$, and $\alpha_0$=15.0\arcdeg\ in the $V$-band and 
$b$=0.0418, $c_1$=0.9149, $c_2$=$-0.9825$, and $\alpha_0$=15.0\arcdeg\ in the $R_\mathrm{C}$-band using the trust--region--reflective algorithm with weighting by the inverse squared error. To search for the marginalized 1-sigma errors of $\alpha_\mathrm{max}$ and $P_\mathrm{max}$, we generated parameter spaces and calculated the chi-square statistic. We thus set the parameter spaces as follows: $ 0.8b_0     < b   < 1.2b_0     $ with $ \Delta b   = 0.01 b_0 $; 
$ 0.4c_{1,0} < c_1 < 1.6c_{1,0} $ with $ \Delta c_1 = 0.01 c_{1,0} $;
and $ 0.7c_{2,0} < c_2 < 1.3c_{2,0} $ with $ \Delta c_2 = 0.01 c_{2,0} $,
where the subscript 0 denotes the ``best fit'' parameter values from the above initial guess and $ \Delta $ means the bin sizes for the parameter search. As a result, we obtained $\alpha_\mathrm{max}$=124$\pm$8\arcdeg\ and  $P_\mathrm{max}$=7.32$\pm$0.25\% in the $V$-band and $\alpha_\mathrm{max}$=124$\pm$6\arcdeg\ and  $P_\mathrm{max}$=7.04$\pm$0.21\% in the $R_\mathrm{C}$-band.

For confirmation, we fit our data with a simple third order polynomial function and obtained $\alpha_\mathrm{max}$=119$\pm$8\arcdeg\ and  $P_\mathrm{max}$=7.26$\pm$0.28\% in the $V$-band and $\alpha_\mathrm{max}$=122$\pm$6\arcdeg\ and  $P_\mathrm{max}$=7.01$\pm$0.13\% in the $R_\mathrm{C}$-band. These results are in good agreement with those with Eq. (\ref{eq:LM}), ensuring reliability of these derived values.


\subsection{Comparison with Other Airless Bodies in the Solar System}
\begin{figure}
\epsscale{0.42}
\plotone{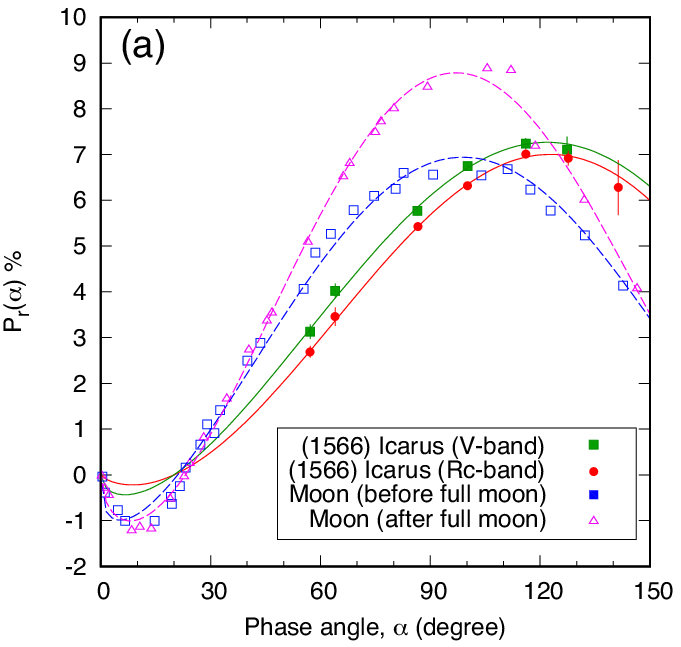}
\plotone{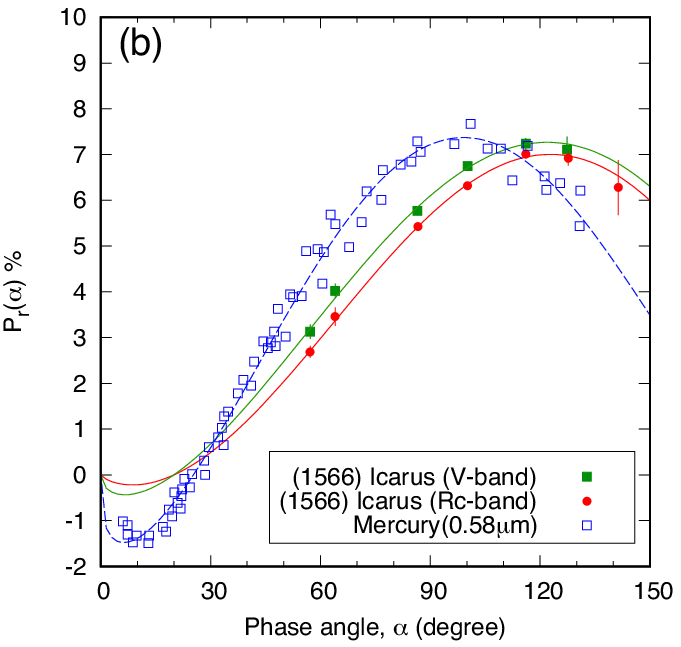}
\plotone{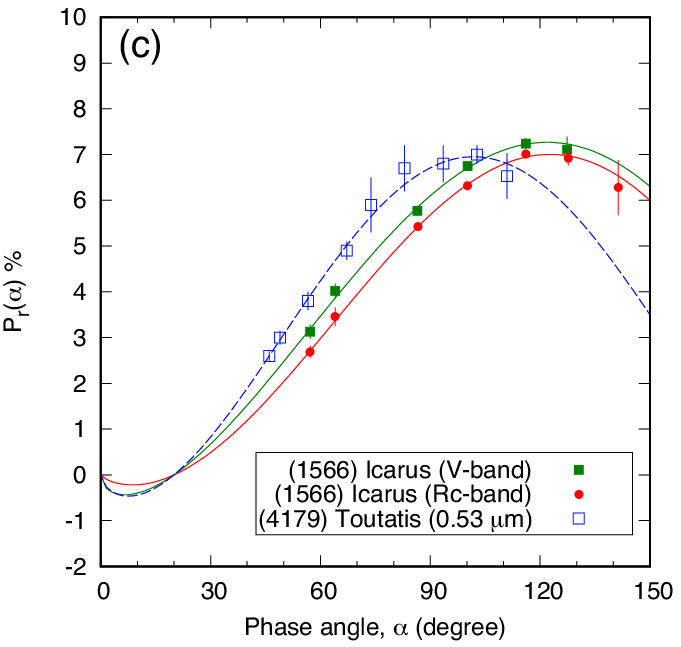}
\plotone{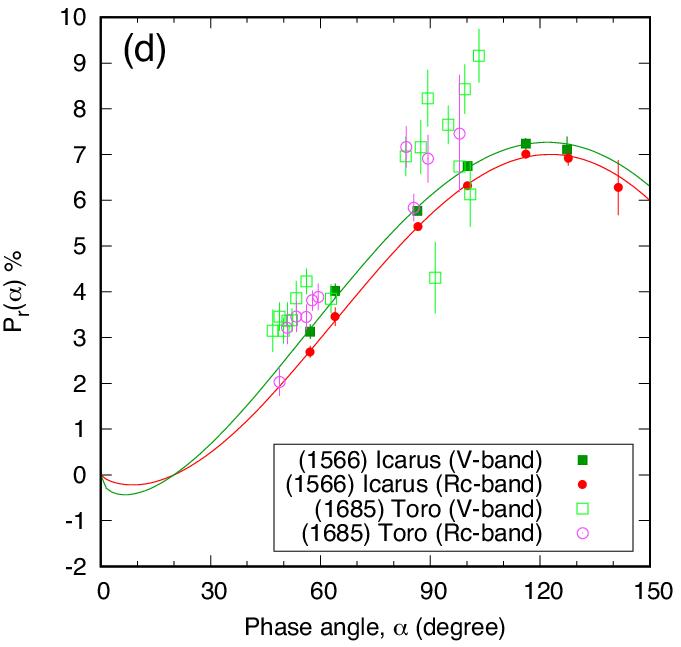}
\plotone{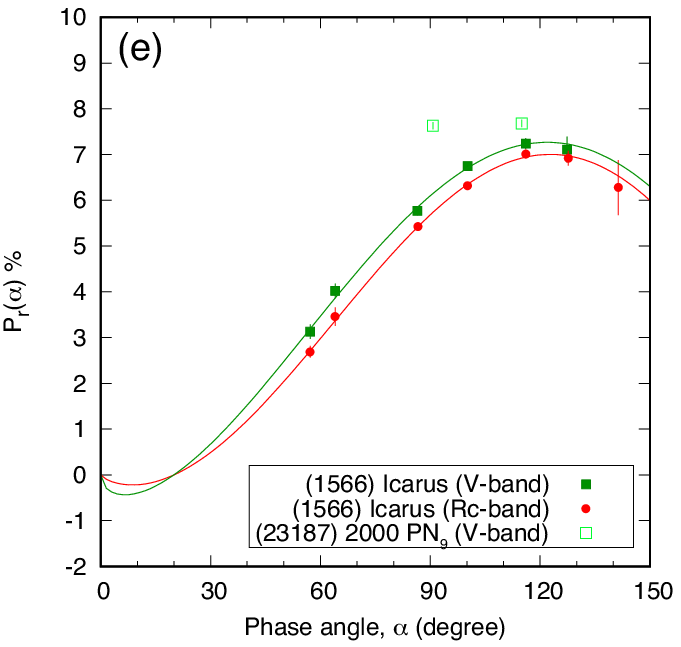}
\plotone{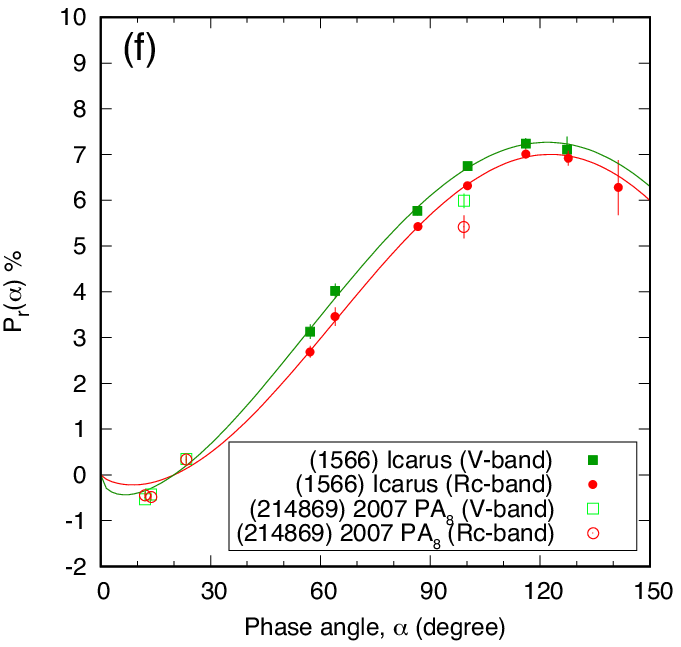}
\caption{Comparison with (a) the Moon \citep{Shkuratov2011}, (b) Mercury \citep{Dollfus1974} and asteroids (c) (4179) Toutatis \citep{Ishiguro1997}, (d) (1685) Toro, (e) (23187) 2000 PN$_9$ \citep{Belskaya2009}, and (f) (214869) 2007 PA$_8$ \citep{Fornasier2015}. The curves fit with Eq. (\ref{eq:LM}) are shown in (a)--(c) but not shown in (d)--(f) because of insufficient phase coverage.
\label{fig:comp}}
\end{figure}

One of the unexpected findings from this research is the large $\alpha_\mathrm{max}$ values. Taking advantage of the observatory's location and the timing (i.e., the summer solstice), we extended the polarimetric data up to $\alpha$=143\arcdeg. The phase angle coverage is overwhelmingly larger than previous polarimetric observations of asteroids, but we marginally detected a drop in polarization beyond $\alpha_\mathrm{max}$ . To make it clear how large $\alpha_\mathrm{max}$ is, we compared the polarimetric phase curves with those of other solar system airless bodies (Figure  \ref{fig:comp}). Among them, the Moon and Mercury are observed well around their maximum polarization \citep[see also,][]{Jeong2015,Dollfus1974}.  Both the Moon and Mercury have $\alpha_\mathrm{max}\sim$ 100\arcdeg, which is significantly smaller than Icarus (see Figure  \ref{fig:comp} (a)--(b)). For asteroids, there are four objects in the literature for which polarization data were available at large $\alpha>$100\arcdeg. \citet{Kiselev1990} made a polarimetric observation of an NEA, Toro, and derived $P_\mathrm{max}$=8.5$\pm$0.7 \% at $\alpha_\mathrm{max}$=110$\pm$10\arcdeg. \citet{Ishiguro1997} conducted an observation of another NEA, Toutatis, at $\alpha$=74\arcdeg--111\arcdeg\ and derived $P_\mathrm{max}$=7.0$\pm$0.2\% at $\alpha_\mathrm{max}$=107$\pm$10\arcdeg. \citet{Belskaya2009} observed 2000 PN$_9$ at $\alpha$=90.7\arcdeg\ and 115\arcdeg\ and posited that $P_\mathrm{max}$=7.7$^{+0.5}_{-0.1}$\% at $\alpha_\mathrm{max}$=103$\pm$12\arcdeg. Among these asteroids, the Toutatis data have good coverage around the maximum phase (Figure \ref{fig:comp} (c)), showing a clear drop beyond $\alpha_\mathrm{max}$. Once again, our Icarus data clearly show $\alpha_\mathrm{max}$ values larger than those for Toutatis.

There are several possibilities resulting in the large $\alpha_\mathrm{max}$. Lunar data show a moderate dependence on albedo \citep{Korokhin2005}. Thus, smaller albedo values tend to show larger $\alpha_\mathrm{max}$ values. Icarus has an albedo typical of stony materials in the solar system (see Section \ref{subsection:albedo}). In addition, lunar data cover an $\alpha_\mathrm{max}$ in the range of 92\arcdeg--106\arcdeg, which is much smaller than the Icarus values. Accordingly, the high $\alpha_\mathrm{max}$ values cannot be explained by the albedo. \citet{Shkuratov1992} examined the size dependence of the polarization properties of laboratory samples and suggested that $\alpha_\mathrm{max}$ would increase with increasing  size, even up to $\alpha_\mathrm{max}\sim$150\arcdeg. Although there may be other factors increasing the $\alpha_\mathrm{max}$ values, we hypothesize that one possible explanation for the large $\alpha_\mathrm{max}$ of Icarus is that the asteroid could be covered with large grains.

\subsection{Albedo}
\label{subsection:albedo}
The geometric albedo ($p_\mathrm{V}$) of Icarus was determined by different measurements, but these results do not match well: 0.42 \citep{Veeder1989}, 0.33--0.70 \citep{Harris2002}, 0.14$^{+0.10}_{-0.06}$ \citep{Thomas2011}, and 0.29$\pm$0.05 \citep{Nugent2015}. We now derive the geometric albedo based on our polarimetric measurement using the so-called slope--albedo law, which is given by

\begin{eqnarray}
\log_{10} p_\mathrm{V} = C_{\rm 1}~\log_{10} h_\mathrm{SLP} + C_2 ~~~~~~,
 \label{eq:slopealbedo}
\end{eqnarray}

\noindent where $h_\mathrm{SLP}$ is the phase slope near the inversion angle (i.e., $dP/d\alpha$ at $\alpha=\alpha_0$). $C_{\rm 1}$ and $C_{\rm 2}$ are constants that have been determined by several authors. \citet{Lupishko1995} derived $C_{\rm 1}$=$-0.98$ and $C_{\rm 2}$=$-1.73$ in an early study, and these values were updated  to $C_{\rm 1}$=$-1.21$ $\pm$ 0.07 and $C_{\rm 2}$=$-1.89$$\pm$$0.14$ \citep{Masiero2012} and to $C_{\rm 1}$=$-0.80$$\pm$$0.04$ and $C_{\rm 2}$=$-1.47$$\pm$$0.04$ (when $p_\mathrm{V}\ge$0.08) \citep{Cellino2015}. We fit our $V$--polarimetric data at $\alpha$=57.2\arcdeg--86.5\arcdeg\, constraining the inversion phase angle $\alpha_0$=20\arcdeg, which is a typical value for Q-type asteroids \citep{Belskaya2017}, and obtained $h_\mathrm{SLP}$=$0.0874$$\pm$$0.0017$. With $C_{\rm 1}$ and $C_{\rm 2}$ as in  \citet{Masiero2012} and \citet{Cellino2015}, we acquired a geometric albedo of $p_\mathrm{V}$=$0.25$$\pm$$0.02$. This albedo value is consistent with those of Q-type and S-type asteroids \citep{Usui2013,DeMeo2013,Thomas2011}. Note that the fitted phase angle range is larger than those of previous studies. However, we believe that this range is reasonable for fitting the data not only because our phase curve shows a linear profile at $\alpha\lesssim$86.5\arcdeg\ but also because a study of Itokawa at similar phase angles of $\alpha$=41.5\arcdeg--79.2\arcdeg\ demonstrated a good match for the albedo ($p_\mathrm{V}$=0.24$\pm$0.01 via polarimetry \citet{Cellino2005} v.s. $p_\mathrm{V}$=0.24$\pm$0.02, \citet{Ishiguro2010}, via remote--sensing observation by the Hayabusa onboard camera).

In Section \ref{subsec:rotation}, we examined the rotational change in $P$ and found no variability to the accuracy of 0.2--0.3\% at $\alpha$=100.2\arcdeg. Extrapolating the linear slope to the phase angle (although the phase curve slightly deviated from the line), the upper limit of the polarization variability (0.2--0.3\%) is converted into the upper limit of the $h_\mathrm{SLP}$ variability of $\sim$0.0025. With Eq. (\ref{eq:slopealbedo}), we put the upper limit of the albedo variation on the quadrant surface at $\sigma p_\mathrm{V}$=0.02. The upper limit would suggest that the surface of the asteroid is quite homogeneous in albedo from a large scale viewpoint  (1/4 of surface resolution).

\subsection{Grain Size Estimate}
\begin{figure}
\epsscale{1.0}
\plotone{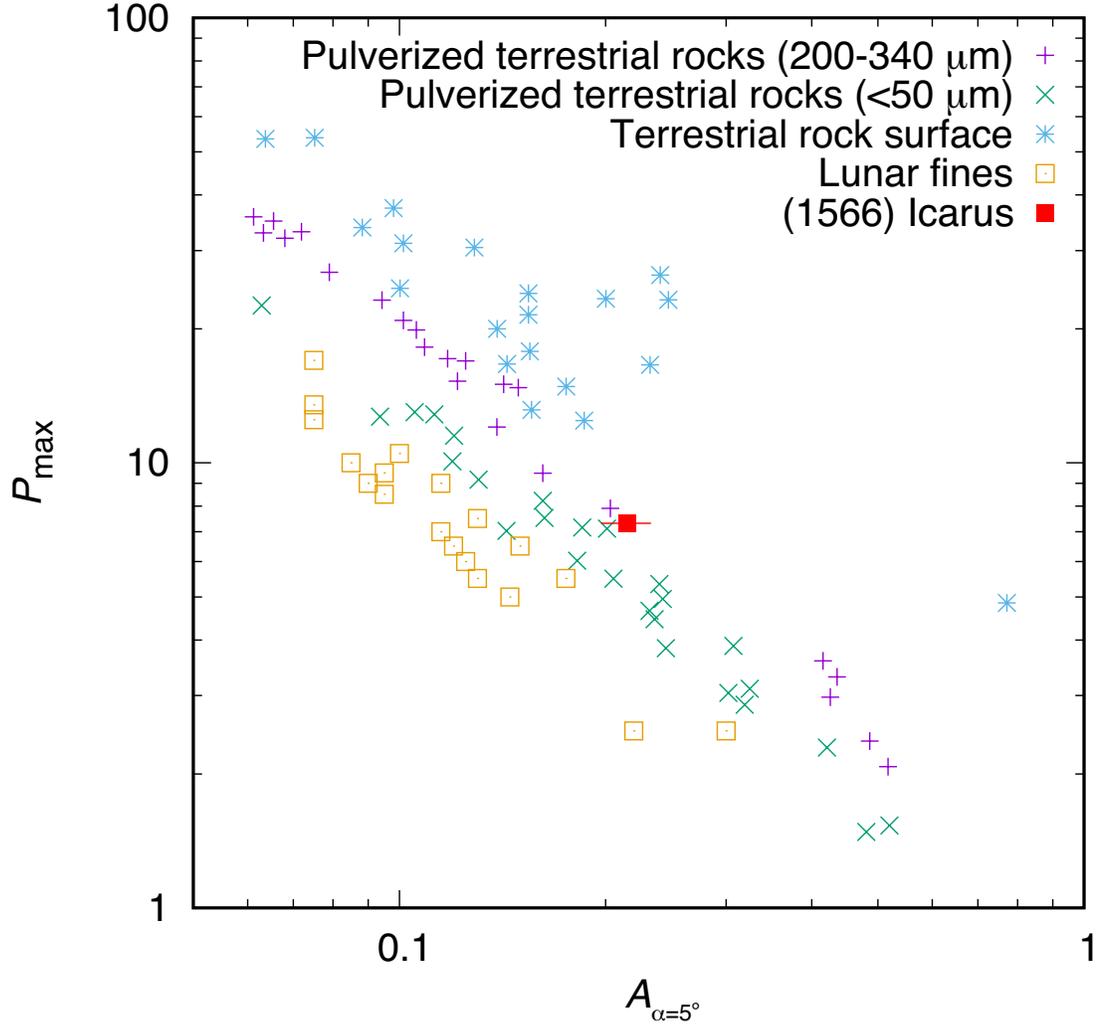}
\caption{Albedo versus $P_\mathrm{max}$ plot for lunar and terrestrial samples in \citep{Geake1986}. We also plotted Icarus data from our measurements. The albedo of these samples ($A_{\alpha=5\arcdeg}$) is defined at a phase angle $\alpha$=5\arcdeg, slightly lower than the geometric albedo ($p_\mathrm{V}$)}.\label{fig:grainsize}
\end{figure}

It is known that $P_\mathrm{max}$ is inversely correlated with the geometric albedo $p_\mathrm{V}$ (Umov law). $P_\mathrm{max}$ also depends on the grain size. \citet{Shkuratov1992} examined these relationship using lunar soil samples and gave the following equations:

\begin{eqnarray}
d = 0.03 \exp(2.9~b)~~,
 \label{eq:so1992a}
\end{eqnarray}

\noindent and

\begin{eqnarray}
b=\log(10^2~A_{\alpha=5\arcdeg})+a\log(10~P_\mathrm{max})~~,
 \label{eq:so1992b}
\end{eqnarray}

\noindent where $d$ denotes the grain size in \micron. $a$ is 0.795 at 0.43 \micron\ and 0.845 at 0.65 \micron\ \citep{Shkuratov1992}. $A_{\alpha=5\arcdeg}$ is the albedo at $\alpha$=5\arcdeg. Using the phase function we determined in Section  \ref{subsec:photometry}, we derived $A_{\alpha=5\arcdeg}$= 0.215$\pm$0.018 for Icarus. Applying Eq. (\ref{eq:so1992b}) to our polarimetric result,  we obtained $d$=100--130 \micron. In addition, we plotted our data onto the $P_\mathrm{max}$--albedo relation for different sizes of laboratory samples (Figure \ref{fig:grainsize}).  Similarly, the plot (Figure \ref{fig:grainsize}) shows a trend indicating that Icarus may be covered with particles hundreds of microns in size.

This result is consistent with the fact that Icarus has a large macroscopic roughness. The large values of $\alpha_\mathrm{max}$ also imply a large particle size. Furthermore, the asteroid exhibits a Q-type spectrum, which is bluer than an S-type spectrum. The blueness can be explained not only by the freshness in terms of the space weathering but also by large grains. It is known that an increase in grain size yields a bluer spectral slope regardless of the types of asteroid \citep[e.g.,][]{Vernazza2016,Reddy2016,Miyamoto1981}. Therefore, these optical properties consistently suggest a large grain size on the asteroid. Why is the particle size so large? How did the asteroid lose the small particles from the surface?

\subsection{Consideration of Mass Ejection around Perihelion}
Icarus has a critical rotational period (2.273 hours) in which the centrifugal force exceeds the self-gravitational force on the equator. Assuming an Itolawa-like bulk density of $\sim$2000 kg m$^{-3}$ \citep{Fujiwara2006,Scheeres2010} and a spherical body with a 1440-m diameter \citep{Greenberg2017}, the ambient gravitational acceleration is approximately 80 micro-G's at the pole and minus 5 micro-G's at the equator, suggesting that granular materials may be ejected from the equatorial region (around a latitude within 30\arcdeg\ from the equator) via centrifugal acceleration. In contrast, the rotational axis of Icarus nearly aligns to the ecliptic pole \citep[][]{Greenberg2017}, meaning that it is roughly perpendicular to the orbital plane with a moderate inclination to the orbital plane ($i$=22.3\arcdeg). Under this geometry, the sun shines almost parallel to the polar region. Although regolith grains can remain in the high-latitude region, the oblique sunshine can strip small grains off from the polar region when the asteroid passes through perihelion. Such an idea was suggested for the surface of (3200) Phaethon to explain the dust emission near perihelion \citep{Jewitt2010,Jewitt2013}.
The solar radiation pressure is given by $F_\mathrm{r}$=$\beta_\mathrm{r} F_\mathrm{g}$, where $F_\mathrm{g}=$0.169 m s$^{-2}$ is the solar gravity at the perihelion of Icarus ($q$=0.187 au). $\beta_\mathrm{r}$ is the ratio of the solar radiation pressure to the solar gravity, given approximately by $\beta_\mathrm{r}$=1.14/$\rho_\mathrm{d} d$, where $\rho_\mathrm{d}$ and $d$ are the particle mass density and diameter, respectively. Thus, the solar radiation pressure exceeds the ambient gravity in the polar region when $d\lesssim$240 \micron\ (a mass density, $\rho_\mathrm{d}$=1.0 g cm$^{-2}$, was assumed). Although some cohesive forces, such as van der Waals forces, would work to prevent mass ejection from the surface, we conjecture that the environment of the ``fast-rotating'' body at a ``small solar distance'' would be responsible for the paucity of small grains and its unique polarimetric properties.

Intriguingly, \citet{Ohtsuka2007} noted that 2007 MK$_6$ has a strong dynamical connection to Icarus, suggesting that these two asteroids share a common origin. Such groupings of asteroids are also recognized for Phaethon and (155140) 2005 UD \citep{Ohtsuka2006}. It is still unclear if these two bodies were split due to the tidal force of a planet during a close encounter, thermal stress, rotational breakup via YORP acceleration, or other mechanisms. It is important to note that these two bodies have similarities in two aspects: their rapid rotational periods and small perihelion distances. Supposing that these groups of asteroids experienced large-scale splittings that produced their current bodies, they may have had the chance to lose small dust grains during splitting due to strong solar radiation pressure quickly sweeping small dust grains from their orbits before they had the chance to accumulate, producing bodies that lack small dust grains.

\section{Summary}
\label{sec:summary}
We made photopolarimetric observations of Icarus at large phase angles $\alpha$= 57\arcdeg--141\arcdeg\ during its apparition in 2015 and found the following:

The combination of the maximum polarization degree and the geometric albedo is in accordance with terrestrial rocks with a diameter of several hundreds of micrometers. The photometric function indicates a large macroscopic roughness.  We posit that the unique environment (i.e., the small perihelion distance $q$=0.187 au and a short rotational period $T_\mathrm{rot}$=2.27 hours) may be attributed to the paucity of small grains on the surface, as indicated on Phaethon.

\begin{enumerate}
\item{ The maximum values of the linear polarization degree are $P_\mathrm{max}$=7.32$\pm$0.25 \% at a phase angle of $\alpha_\mathrm{max}$=124\arcdeg$\pm$8\arcdeg\ in the $V$-band and $P_\mathrm{max}$=7.04$\pm$0.21 \% at $\alpha_\mathrm{max}$=124\arcdeg$\pm$6\arcdeg\ in the $R_\mathrm{C}$--band.}
\item{Applying the polarimetric slope--albedo law, we derived a geometric albedo $p_\mathrm{V}$=0.25$\pm$0.02, which is consistent with that of Q-type asteroids. The albedo would be globally constant, showing no significant rotational variation in the polarization degree.}
\item{$\alpha_\mathrm{max}$ is significantly larger than those of Mercury, the Moon and the S--type asteroid Toutatis but consistent with laboratory samples hundreds of microns in size.}
\item{The $P_\mathrm{max}$--albedo relation suggests that Icarus is covered with particles hundreds of microns in size.}
\item{The photometric function suggest a large macroscopic roughness, supporting the dominance of large grains.}
\end{enumerate}

To explain the dominance of large grains on the asteroid, we conjecture that a strong radiation pressure around the perihelion passage would strip small grains off of the fast--rotating asteroid.

\newpage

\appendix
\label{sec:appendix1}
\section{Pirka/MSI Polarimetric Data Analysis Procedures}

The observed ordinary and extraordinary fluxes at the half-wave plate angle $\Psi$ in degrees, $I_\mathrm{o}(\Psi)$ and $I_\mathrm{e}(\Psi)$, were used to derive

\begin{eqnarray}
q_\mathrm{pol}'=\left(\frac{R_\mathrm{q}-1}{R_\mathrm{q}+1}\right)\bigg/p_\mathrm{eff}~,
\label{eq:Q}
\end{eqnarray}

\noindent and

\begin{eqnarray}
u_\mathrm{pol}'=\left(\frac{R_\mathrm{u}-1}{R_\mathrm{u}+1}\right)\bigg/p_\mathrm{eff}~,
\label{eq:U}
\end{eqnarray}

\noindent where $R_\mathrm{q}$ and $R_\mathrm{u}$ are obtained from the observation using the following equations:

\begin{eqnarray}
R_\mathrm{q}=\sqrt{\frac{I_\mathrm{e}(0)/I_\mathrm{o}(0)}{I_\mathrm{e}(45)/I_\mathrm{o}(45)}}~,
\label{eq:Rq}
\end{eqnarray}

\noindent and

\begin{eqnarray}
R_\mathrm{u}=\sqrt{\frac{I_\mathrm{e}(22.5)/I_\mathrm{o}(22.5)}{I_\mathrm{e}(67.5)/I_\mathrm{o}(67.5)}}~,
\label{eq:Ru}
\end{eqnarray}

\noindent where $p_\mathrm{eff}$ is a polarization efficiency, which was examined by taking a dome flat image through a pinhole and a
Polaroid--like linear polarizer, which produces artificial stars with $P$=99.97$\pm$0.02 \% ($V$) and 99.98$\pm$0.01 \% ($R_\mathrm{C}$). $p_\mathrm{eff}$ was measured
approximately two months prior to our observation and was determined to be $p_\mathrm{eff}$=0.9967$\pm$0.0003 in the $V$-band and 0.9971$\pm$0.0001 in the $R_\mathrm{C}$-band.

The instrumental polarization of Pirka/MSI is known to depend on the instrument angle of rotation and can be corrected with the following equation:

\begin{eqnarray}
\left( \begin{array}{r} q''_\mathrm{pol} \\ u''_\mathrm{pol} \end{array} \right)
 = \left( \begin{array}{r} q'_\mathrm{pol} \\ u'_\mathrm{pol} \end{array} \right)
 	- \left( \begin{array}{rr} \cos 2\theta_\mathrm{rot1} & -\sin 2\theta_\mathrm{rot1} \\ \sin 2\theta_\mathrm{rot2} & \cos 2\theta_\mathrm{rot2} \end{array} \right)
	\left( \begin{array}{r} q_\mathrm{inst} \\ u_\mathrm{inst} \end{array} \right)~,
\label{eq:instpol}
\end{eqnarray}

\noindent where $\theta_\mathrm{rot1}$ denotes the average instrument rotator angle during the exposures with $\Psi$=0\arcdeg\ and 45\arcdeg, while $\theta_\mathrm{rot2}$ denotes the average angle with $\Psi$=22.5\arcdeg\ and 67.5\arcdeg. $q_\mathrm{inst}$ and $u_\mathrm{inst}$ are two components of the Stokes parameters for the instrumental polarization and were determined to be $q_\mathrm{inst}$=0.963$\pm$0.029 \% in the $V$-band and 0.703$\pm$0.033 \% in the $R_\mathrm{C}$-band and $u_\mathrm{inst}$=0.453$\pm$0.043 \% in the $V$-band and 0.337$\pm$0.020 \% in the $R_\mathrm{C}$-band, respectively, by observing the unpolarized stars HD212311 and BD+32 3739 \citep[see Table 3, on page 1566, ][]{Schmidt1992}. 

The instrument position angle in celestial coordinates was determined by measuring the polarization position angles of strongly
polarized stars for which position angles are reported in \citet{Schmidt1992}. The instrument position angle can be corrected using the following equations:

\begin{eqnarray}
\left( \begin{array}{r} q'''_\mathrm{pol} \\ u'''_\mathrm{pol} \end{array} \right)
 =	\left( \begin{array}{rr} \cos 2\theta'_\mathrm{off} & \sin 2\theta'_\mathrm{off} \\ -\sin 2\theta'_\mathrm{off} & \cos 2\theta'_\mathrm{off} \end{array} \right)
	\left( \begin{array}{r} q''_\mathrm{pol} \\ u''_\mathrm{pol} \end{array} \right)~~ ,
\label{eq:thetaoffset}
\end{eqnarray}

\noindent and

\begin{eqnarray}
\theta'_\mathrm{off} = \theta_\mathrm{off} - \theta_\mathrm{ref}~~,
\label{eq:thetaoffset2}
\end{eqnarray}

\noindent where  $\theta_\mathrm{ref}$ is a given parameter for specifying the position angle of the instrument. Through an observation of strongly polarized stars (HD204827, HD154445, and HD155197) in 2015 May, we derived $\theta_\mathrm{off}$=3.82$\pm$0.38\arcdeg\ in the V-band and 3.38$\pm$0.37 \arcdeg\ in the $R_\mathrm{C}$-band.

\newpage

\vspace{1cm}
{\bf Acknowledgments}\\
This research was supported by a National Research Foundation of Korea (NRF) grant funded by the Korean government (MEST) (No. 2015R1D1A1A01060025). The observations at the Nayoro Observatory were supported by the Optical and Near-infrared Astronomy Inter-University Cooperation Program and Grants-in-Aid for Scientific Research (23340048, 24000004, 24244014, and 24840031) from the Ministry of Education, Culture, Sports, Science and Technology of Japan. We thank the staff members at the Nayoro City Observatory, Y. Murakami, F. Watanabe, Y. Kato, and R. Nagayoshi, for their kind support and Drs. Takashi Ito and Tomoko Arai for their encouragement regarding this work.
We also thank the anonymous reviewer for their careful reading of our manuscript and their insightful comments. SH was supported by the Hypervelocity Impact Facility (former name: the Space Plasma Laboratory), ISAS, JAXA.

\begin{deluxetable}{cccrrrrrrrr} 
\tablecolumns{11} 
\tablewidth{0pc} 
\tablecaption{Observation Journal.} 
\tablehead{ 
\colhead{Date} & \colhead{UT} & \colhead{Filter}  & \colhead{$t_\mathrm{exp}^{(a)}$} & \colhead{$N^{(b)}$} & \colhead{r$^{(c)}$} & \colhead{$\Delta^{(d)}$} & \colhead{$\alpha^{(e)}$} & \colhead{$\theta_\perp^{(f)}$} &  \colhead{$\phi^{(g)}$} & \colhead{mode$^{(h)}$}
}
\startdata
2015/06/11&  11:23-15:27 &R$_\mathrm{C}$& 60 & 212 & 0.928 & 0.104 & 145.1 & 86.3 &356.3 & Phot\\
2015/06/12&  13:22-16:18 &R$_\mathrm{C}$& 60 & 92 & 0.945 & 0.089 & 141.3 & 98.9 & 8.9 & Phot, Pol\\
2015/06/14&  13:48-15:50 &V& 60 & 36 & 0.975 & 0.064 & 127.4 & 144.7 & 54.7 & Phot, Pol\\
 &  13:10-15:09 &R$_\mathrm{C}$& 60 & 48 & 0.975 & 0.065 & 127.7 & 143.7 & 53.7 & Pol\\
2015/06/15&  11:34-15:27 &V& 30 & 152 & 0.989 & 0.057 & 116.1 & 176.0 & 86.0 & Pol\\
 &  11:42-15:18 &R$_\mathrm{C}$& 30 & 156 & 0.989 & 0.057 & 116.1 & 176.0 & 86.0 & Pol\\
2015/06/16&  12:30-17:20 &V& 30 & 220 & 1.005 & 0.054 & 100.2 & 20.6 & 110.6 & Pol\\
 &  12:22-17:13 &R$_\mathrm{C}$& 30 & 200 & 1.005 & 0.054 & 100.2 & 20.5 & 110.5 & Pol\\
2015/06/17&  11:19-12:43 &V& 30 & 68 & 1.018 & 0.056 & 86.5 & 29.8 & 119.8 & Pol\\
 &  11:08-12:35 &R$_\mathrm{C}$& 30 & 76 & 1.018 & 0.056 & 86.6 & 29.8 & 119.8 & Pol\\
2015/06/19&  11:13-11:20 &V& 30 & 12 & 1.046 & 0.073 & 64.0 & 32.2 & 122.2 & Pol\\
 &  11:02-11:12 &R$_\mathrm{C}$& 30 & 16 & 1.046 & 0.073 & 64.0 & 32.2 & 122.2 & Pol\\
2015/06/20&  11:22-11:29 &V& 30 & 12 & 1.060 & 0.086 & 57.2 & 30.1 & 120.1 & Pol\\
 &  11:12-11:22 &R$_\mathrm{C}$& 30 & 16 & 1.060 & 0.086 & 57.2 & 30.1& 120.1 & Pol\\
\enddata 
   \tablenotetext{(a)}{Individual effective exposure time in seconds.}
   \tablenotetext{(b)}{Number of exposures.}
   \tablenotetext{(c)}{Median heliocentric  distance in au.}
   \tablenotetext{(d)}{Median geocentric  distance in au.}
   \tablenotetext{(e)}{Median Solar phase angle (Sun--Asteroid--Observer angle) in degrees.}
   \tablenotetext{(f)}{Median position angle of normal vector with respect to the scattering plane in degrees.}
   \tablenotetext{(g)}{Median position angle of the scattering plane in degrees.}
   \tablenotetext{(h)}{Observation mode: Photometry (Phot) or Polarimetry (Pol).}
\label{tab:t1}
\end{deluxetable} 
\clearpage

\begin{deluxetable}{ccrrrrrr} 
\tablecolumns{8} 
\tablewidth{0pc} 
\tablecaption{Degree of linear polarization and position angle of polarization} 
\tablehead{ 
\colhead{Date} & \colhead{Filter}  & \colhead{$P$$^{(a)}$}  & \colhead{$\epsilon P$$^{(b)}$} & \colhead{$\theta_\mathrm{P}^{(c)}$} 
& \colhead{$\epsilon \theta_\mathrm{P}^{(d)}$} & \colhead{$P_\mathrm{r}$$^{(e)}$} &  \colhead{$\theta_\mathrm{r}$$^{(f)}$}
}
\startdata
2015/06/12 & R$_\mathrm{C}$ & 6.29  & 0.60  &  $-82.8$  & 2.7 & 6.28 & $-1.7$\\
2015/06/14 & V & 7.14  & 0.28  &  $-37.8$  & 1.1 & 7.11 & $-2.5$\\
 & R$_\mathrm{C}$ & 6.92  & 0.16  &  $-36.7$  & 0.7 & 6.92 & $-0.4$\\
2015/06/15 & V & 7.26  & 0.12  &  $-5.8$  & 0.5 & 7.24 & $-1.8$\\
 & R$_\mathrm{C}$ & 7.02  & 0.09  &  $-5.3$  & 0.4 & 7.01 & $-1.3$\\
2015/06/16 & V & 6.77  & 0.08  &  18.4  & 0.3 & 6.75 & $-2.2$\\
 & R$_\mathrm{C}$ & 6.33  & 0.06  &  18.8  & 0.3 & 6.32 & $-1.7$\\
2015/06/17 & V & 5.78  & 0.09  &  28.2  & 0.4 & 5.77 & $-1.6$\\
 & R$_\mathrm{C}$ & 5.44  & 0.07  &  28.2  & 0.4 & 5.43 & $-1.6$\\
2015/06/19 & V & 4.02  & 0.16  &  31.5  & 1.1 & 4.02 & $-0.7$\\
 & R$_\mathrm{C}$ & 3.47  & 0.20  &  34.6  & 1.7 & 3.46 & $2.4$\\
2015/06/20 & V & 3.15  & 0.16  &  27.3  & 1.5 & 3.13 & $-2.8$\\
 & R$_\mathrm{C}$ & 2.71  & 0.13  &  27.1  & 1.4 & 2.69 & $-3.0$\\
\enddata 
\tablenotetext{(a)}{Polarization degree in percent.}
\tablenotetext{(b)}{Error of $P$ in percent.}
\tablenotetext{(c)}{Position angle of the strongest electric vector in degrees.}
\tablenotetext{(d)}{Error of $\theta_P$ in degrees.}
\tablenotetext{(e)}{Polarization degree with respect to the scattering plane in percent.}
\tablenotetext{(f)}{Position angle with respect to the scattering plane in degrees.}

\label{polresult}
\end{deluxetable} 

\clearpage

\clearpage

\clearpage

\clearpage

\clearpage

\clearpage

\clearpage


\newpage
\begin{thebibliography}{}
\bibitem[Allen(1973)]{Allen1973} Allen, C.~W.\ 1973, London: University of London, Athlone Press, |c1973, 3rd ed.,  
\bibitem[Belskaya et al.(2009)]{Belskaya2009} Belskaya, I.~N., Levasseur-Regourd, A.-C., Cellino, A., et al.\ 2009, \icarus, 199, 97 
\bibitem[Belskaya et al.(2017)]{Belskaya2017} Belskaya, I.~N., Fornasier, S., Tozzi, G.~P., et al.\ 2017, \icarus, 284, 30 
\bibitem[Bowell et al.(1972)]{Bowell1972} Bowell, E., Dollfus, A., \& Geake, J.~E.\ 1972, Lunar and Planetary Science Conference Proceedings, 3, 3103 
\bibitem[Cellino et al.(2016)]{Cellino2016} Cellino, A., Bagnulo, S., Gil-Hutton, R., et al.\ 2016, \mnras, 455, 2091 
\bibitem[Cellino et al.(2015)]{Cellino2015} Cellino, A., Bagnulo, S., Gil-Hutton, R., et al.\ 2015, \mnras, 451, 3473 
\bibitem[Cellino et al.(2005)]{Cellino2005} Cellino, A., Yoshida, F., Anderlucci, E., et al.\ 2005, \icarus, 179, 297 
\bibitem[Chapman et al.(1975)]{Chapman1975} Chapman, C.~R., Morrison, D., \& Zellner, B.\ 1975, \icarus, 25, 104 
\bibitem[DeMeo et al.(2014)]{DeMeo2014} DeMeo, F.~E., Binzel, R.~P., \& Lockhart, M.\ 2014, \icarus, 227, 112 
\bibitem[DeMeo \& Carry(2013)]{DeMeo2013} DeMeo, F.~E., \& Carry, B.\ 2013, \icarus, 226, 723 
\bibitem[van Dokkum(2001)]{Dokkum2001} van Dokkum, P.~G.\ 2001, \pasp, 113, 1420 
\bibitem[Dollfus(1998)]{Dollfus1998} Dollfus, A.\ 1998, \icarus, 136, 69 
\bibitem[Dollfus et al.(1989)]{Dollfus1989} Dollfus, A., Wolff, M., Geake, J.~E., Dougherty, L.~M., \& Lupishko, D.~F.\ 1989, Asteroids II, 594 
\bibitem[Dollfus \& Auriere(1974)]{Dollfus1974} Dollfus, A., \& Auriere, M.\ 1974, \icarus, 23, 465 
\bibitem[Fujiwara et al.(2006)]{Fujiwara2006} Fujiwara, A., Kawaguchi, J., Yeomans, D.~K., et al.\ 2006, Science, 312, 1330 
\bibitem[Fornasier et al.(2015)]{Fornasier2015} Fornasier, S., Belskaya, I.~N., \& Perna, D.\ 2015, \icarus, 250, 280 
\bibitem[Geake \& Dollfus(1986)]{Geake1986} Geake, J.~E., \& Dollfus, A.\ 1986, \mnras, 218, 75 
\bibitem[Gehrels et al.(1970)]{Gehrels1970} Gehrels, T., Roemer, E., Taylor, R.~C., \& Zellner, B.~H.\ 1970, \aj, 75, 186 
\bibitem[Gil-Hutton et al.(2014)]{GH2014} Gil-Hutton, R., Cellino, A., \& Bendjoya, P.\ 2014, \aap, 569, A122 
\bibitem[Gil-Hutton \& Ca{\~n}ada-Assandri(2012)]{GH2012} Gil-Hutton, R., \& Ca{\~n}ada-Assandri, M.\ 2012, \aap, 539, A115 
\bibitem[Gil-Hutton \& Ca{\~n}ada-Assandri(2011)]{GH2011} Gil-Hutton, R., \& Ca{\~n}ada-Assandri, M.\ 2011, \aap, 529, A86 
\bibitem[Goidet-Devel et al.(1995)]{Goidet-Devel1995} Goidet-Devel, B., Renard, J.~B., \& Levasseur-Regourd, A.~C.\ 1995, \planss, 43, 779
\bibitem[Granvik et al.(2016)]{Granvik2016} Granvik, M., Morbidelli, A., Jedicke, R., et al.\ 2016, \nat, 530, 303 
\bibitem[Greenberg et al.(2017)]{Greenberg2017} Greenberg, A.~H., Margot, J.-L., Verma, A.~K., et al.\ 2017, \aj, 153, 108 
\bibitem[Hapke(1984)]{Hapke1984} Hapke, B.\ 1984, Icarus, 59, 41 
\bibitem[Hapke(1993)]{Hapke1993} Hapke, B.\ 1993, Topics in Remote, Sensing, Cambridge, UK: Cambridge University Press
\bibitem[Harris(1998)]{Harris1998} Harris, A.~W.\ 1998, \icarus, 131, 291 
\bibitem[Harris \& Lagerros(2002)]{Harris2002} Harris, A.~W., \& Lagerros, J.~S.~V.\ 2002, Asteroids III, W.~F.~Bottke Jr., A.~Cellino, P.~Paolicchi, and R.~P.~Binzel (eds), University of Arizona Press, Tucson, p.205-218, 205 
\bibitem[Helfenstein et al.(1996)]{Helfenstein1996} Helfenstein, P., et  al.\ 1996, Icarus, 120, 48 
\bibitem[Helfenstein et al.(1994)]{Helfenstein1994} Helfenstein, P., et  al.\ 1994, Icarus, 107, 37 
\bibitem[Itoh et al.(2017)]{Itoh2016} Itoh, R., Tanaka, Y.~T., Kawabata, K.~S., et al.\ 2017, \pasj, 69, 25
\bibitem[Ishiguro et al.(1997)]{Ishiguro1997} Ishiguro, M., Nakayama, H., Kogachi, M., et al.\ 1997, \pasj, 49, L31 
\bibitem[Ishiguro et al.(2010)]{Ishiguro2010} Ishiguro, M., Nakamura, R., Tholen, D.~J., et al.\ 2010, \icarus, 207, 714 
\bibitem[Jeong et al.(2015)]{Jeong2015} Jeong, M., Kim, S.~S., Garrick-Bethell, I., et al.\ 2015, \apjs, 221, 16
\bibitem[Jewitt \& Li(2010)]{Jewitt2010} Jewitt, D., \& Li, J.\ 2010, \aj, 140, 1519 
\bibitem[Jewitt et al.(2013)]{Jewitt2013} Jewitt, D., Li, J., \& Agarwal, J.\ 2013, \apjl, 771, L36 
\bibitem[Jewitt(2013)]{Jewitt2013b} Jewitt, D.\ 2013, \aj, 145, 133 
\bibitem[Kiselev et al.(1990)]{Kiselev1990} Kiselev, N.~N., Lupishko, D.~F., Chernova, G.~P., \& Shkuratov, I.~G.\ 1990, Kinematika i Fizika Nebesnykh Tel, 6, 77 
\bibitem[Korokhin \& Velikodsky(2005)]{Korokhin2005} Korokhin, V.~V., \& Velikodsky, Y.~I.\ 2005, Solar System Research, 39, 45 
\bibitem[Kuroda et al.(2015)]{Kuroda2015} Kuroda, D., Ishiguro, M., Watanabe, M., et al.\ 2015, \apj, 814, 156
\bibitem[Lederer et al.(2008)]{Lederer2008} Lederer, S.~M., Domingue, D.~L., Thomas-Osip, J.~E., Vilas, F., Osip, D.~J., Leeds, S.~L., \& Jarvis, K.~S.\ 2008, Earth, Planets, and Space, 60, 49  
\bibitem[Lupishko et al.(1995)]{Lupishko1995} Lupishko, D.~F., Vasilyev, S.~V., Efimov, J.~S., \& Shakhovskoj, N.~M.\ 1995, \icarus, 113, 200 
\bibitem[Mahapatra et al.(1999)]{Mahapatra1999} Mahapatra, P.~R., Ostro, S.~J., Benner, L.~A.~m., et al.\ 1999, \planss, 47, 987
\bibitem[Masiero et al.(2012)]{Masiero2012} Masiero, J.~R., Mainzer, A.~K., Grav, T., et al.\ 2012, \apj, 749, 104 
\bibitem[Miner \& Young(1969)]{Miner1969} Miner, E., \& Young, J.\ 1969, \icarus, 10, 436 
\bibitem[Miyamoto et al.(1981)]{Miyamoto1981} Miyamoto, M., Mito, A., Takano, Y., \& Fujii, N.\ 1981, National Institute Polar Research Memoirs, 20, 345 
\bibitem[Mukai et al.(1997)]{Mukai1997} Mukai, T., Iwata, T., Kikuchi, S., et al.\ 1997, \icarus, 127, 452 
\bibitem[Nakayama et al.(2000)]{Nakayama2000} Nakayama, H., Fujii, Y., Ishiguro, M., et al.\ 2000, \icarus, 146, 220 
\bibitem[Nugent et al.(2015)]{Nugent2015} Nugent, C.~R., Mainzer, A., Masiero, J., et al.\ 2015, \apj, 814, 117 
\bibitem[Ohtsuka et al.(2007)]{Ohtsuka2007} Ohtsuka, K., Arakida, H., Ito, T., et al.\ 2007, \apjl, 668, L71 
\bibitem[Ohtsuka et al.(2006)]{Ohtsuka2006} Ohtsuka, K., Sekiguchi, T., Kinoshita, D., et al.\ 2006, \aap, 450, L25 
\bibitem[Penttil\"{a} et al.(2005)]{Penttila2005} Penttil\"{a}, A., Lumme, K., Hadamcik, E., et al.\ 2005, \aap, 432, 1081
\bibitem[Reddy et al.(2016)]{Reddy2016} Reddy, V., Sanchez, J.~A., Bottke, W.~F., et al.\ 2016, \aj, 152, 162 
\bibitem[Scheeres et al.(2010)]{Scheeres2010} Scheeres, D.~J., Hartzell, C.~M., S{\'a}nchez, P., \& Swift, M.\ 2010, \icarus, 210, 968 
\bibitem[Schmidt et al.(1992)]{Schmidt1992} Schmidt, G.~D., Elston, R., \& Lupie, O.~L.\ 1992, \aj, 104, 1563
\bibitem[Shkuratov \& Opanasenko(1992)]{Shkuratov1992} Shkuratov, I.~G., \& Opanasenko, N.~V.\ 1992, \icarus, 99, 468
\bibitem[Shkuratov et al.(2011)]{Shkuratov2011} Shkuratov, Y., Kaydash, V., Korokhin, V., et al.\ 2011, \planss, 59, 1326 
\bibitem[Takahashi et al.(2009)]{Takahashi2009} Takahashi, S., Yoshida, F., Shinokawa, K., Mukai, T., \& Kawabata, K.~S.\ 2009, \aj, 138, 951 
\bibitem[Takahashi et al.(2004)]{Takahashi2004} Takahashi, S., Shinokawa, K., Yoshida, F., et al.\ 2004, Earth, Planets, and Space, 56, 997 
\bibitem[Thomas et al.(2011)]{Thomas2011} Thomas, C.~A., Trilling, D.~E., Emery, J.~P., et al.\ 2011, \aj, 142, 85
\bibitem[Usui et al.(2013)]{Usui2013} Usui, F., Kasuga, T., Hasegawa, S., et al.\ 2013, \apj, 762, 56 
\bibitem[Veeder et al.(1989)]{Veeder1989} Veeder, G.~J., Hanner, M.~S., Matson, D.~L., et al.\ 1989, \aj, 97, 1211 
\bibitem[Vernazza et al.(2016)]{Vernazza2016} Vernazza, P., Marsset, M., Beck, P., et al.\ 2016, \aj, 152, 54 
\bibitem[Veverka et al.(1988)]{Veverka1988} Veverka, J., Helfenstein, P., Hapke, B., \& Goguen, J.~D.\ 1988, Mercury, University of Arizona Press, 37 
\bibitem[Warner(2015)]{Warner2015} Warner, B.~D.\ 2015, Minor Planet Bulletin, 42, 256 
\bibitem[Watanabe et al.(2012)]{Watanabe2012} Watanabe, M., Takahashi, Y., Sato, M., et al.\ 2012, \procspie, 8446, 84462O 
\bibitem[Zacharias et al.(2010)]{Zacharias2009} Zacharias, N., Finch, C., Girard, T., et al.\ 2010, \aj, 139, 2184


\end{thebibliography}
\end{document}